\documentclass[numberedheadings]{aipproc}
\layoutstyle{6x9}
\usepackage{amsmath,graphicx,makeidx,mathrsfs}
\begin{document}
%%%%%%%%%%%%%%%%%%%%%%%%%%%%%%
\begin{flushright}
\begin{tabular}{l}
  ADP-03-124/T561 \\
  PCCF-RI-03-0308 \\
\end{tabular}  
\end{flushright}
%%%%%%%%%%%%%%%%%%%%%%%%%%%%%%%
 
\title{QCD Factorization in $B$ Decays into $\rho \pi$\footnote{Presented at Fourth Tropical Workshop,
Cairns, Australia, 9-13 June 2003.}}
\author{X.-H. Guo$^\dagger$\footnote{xhguo@physics.adelaide.edu.au}\ , 
\underline {O.M.A. Leitner}$^{\dagger,\ddagger}$\footnote{oleitner@physics.adelaide.edu.au}\ , 
A.W. Thomas$^\dagger$\footnote{athomas@physics.adelaide.edu.au}}
{address={$^\dagger$ Department of Physics  and \\
Special Research Centre for the Subatomic Structure of Matter, \\
University of Adelaide, Adelaide 5005, Australia \\
$^\ddagger$ Laboratoire de Physique Corpusculaire de Clermont-Ferrand \\
IN2P3/CNRS Universit\'e Blaise Pascal \\
F-63177 Aubi\`ere Cedex France }}
\begin{abstract}
Based on the QCD factorization approach we analyse the branching ratios 
for the channel $B \to \rho  \pi$. From the comparisons with
experimental data provided by CLEO, BELLE and BABAR we  constrain the form 
factor $F^{B \to \pi}(m_{\rho}^{2})$ and  propose  boundaries  for this form factor depending on the 
CKM matrix element parameters $\rho$ and $\eta$.
\end{abstract}
\maketitle
%
%#########################################################################################
%-----------------------------------------------------------------------------------------
\section{Naive factorization}
%-----------------------------------------------------------------------------------------
%#########################################################################################
%
The investigation of $B$ decays requires a knowledge of both the  soft and hard interactions which control the 
dynamics   of quarks and gluons. Because  the energy involved in  $B$ decays 
 covers a large range,  from $m_{b}$ down to $\Lambda_{QCD}$, it is necessary to describe the
phenomenon with accuracy. Recently, the BELLE, BABAR, and CLEO facilities have been providing more and more data 
which can be compared with  theoretical results and hence increase   our  knowledge  in  this area.

In any phenomenological treatment of the weak decays of hadrons, the starting point is the weak effective
Hamiltonian at low energy~\cite{Buras:1999rb, Buras:1998ra, Buchalla:1996vs, Stech:1997ij, Buras:1995iy}. It is
obtained by integrating out the heavy fields (e.g. the top quark, $W$ and $Z$ bosons) from the Standard Model
 Lagrangian. It can be written as,
\begin{equation}\label{eq3.9}
{\cal H}_{eff}=\frac {G_{F}}{\sqrt 2} \sum_{i} V_{CKM} C_{i}(\mu)O_i(\mu)\ ,
\end{equation}
where $G_{F}$ is the Fermi constant, $V_{CKM}$ is the CKM matrix element, 
$C_{i}(\mu)$ are the Wilson coefficients, $O_i(\mu)$ are the operators entering 
the Operator Product Expansion  and $\mu$ represents the renormalization scale.
 In the present case, since we analyse direct $CP$ violation in $B$ decays into $\rho \pi$,
 we take into account both tree and penguin operators and  the effective Hamiltonian is,
\begin{equation}\label{eq3.10}
{\cal H}_{eff}^{\bigtriangleup B=1}=\frac {G_{F}}{\sqrt 2} \Biggl[  V_{ub}V_{uq}^{*}\bigl(C_{1}O_{1}^{q} +
C_{2}O_{2}^{q}\bigr)- V_{tb}V_{tq}^{*} \sum_{i=3}^{10} C_{i}O_{i} \Biggr] + h.c.\ ,
\end{equation}
%
%\vspace{-0.5em}
where $q=d$. Consequently, the decay amplitude can  be expressed as  follows,
\begin{multline}\label{eq3.11}
A(B \rightarrow P V) =
\frac {G_{F}}{\sqrt 2} \Biggl[ V_{ub}V_{uq}^{*}\biggl( C_{1}\langle P V | O_{1}^{q}| B \rangle +
C_{2}\langle P V |O_{2}^{q}| B \rangle \biggr) - \\
 V_{tb}V_{tq}^{*} \sum_{i=3}^{10} C_{i}\langle P V |O_{i}| B \rangle \Biggr]+ h.c.\ ,
\end{multline}
where $\langle P V |O_{i}| B \rangle$ are the hadronic matrix elements, and   $P(V)$ indicates  a
pseudoscalar (vector) meson. The matrix elements  describe the transition between initial
  and final state at scales lower than $\mu$ and include, up to now, the main  uncertainties in the calculation
because  it involves  the non-perturbative physics.

\begin{figure}[htbp]
\includegraphics*[width=0.695\columnwidth]{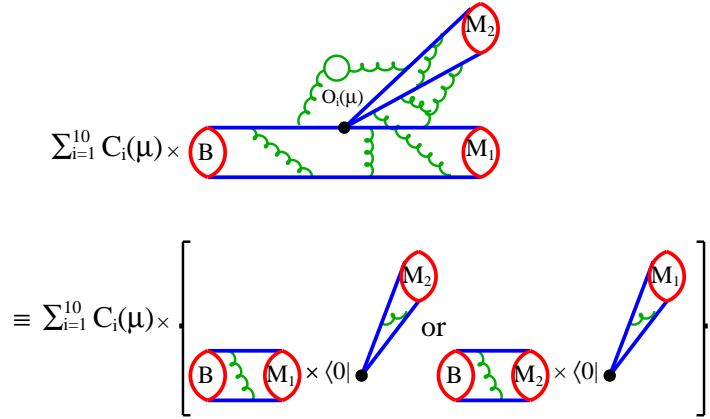}
\caption{Naive factorization, where $M_{1}$ and $M_{2}$ represent the final meson  states.}
\label{figch3.5}
\end{figure}

The computation of  the hadronic matrix elements, $\langle P V |O_{i}| B \rangle$, is not  trivial and requires some
assumptions. The general method which has been used is the so-called  ``factorization'' procedure~\cite{Fakirov:1978ta,
Cabibbo:1978zv, Dugan:1991de}, 
in which one  approximates  the matrix element as a product of a transition matrix element
between a $B$ meson and one final state meson and  a matrix element which describes the creation of 
the second meson from the vacuum. This can be formulated as,
\begin{align}\label{eq3.12}
\langle P V |O_{i}| B \rangle =& \langle V  | J_{2i}| 0 \rangle\ \langle P  | J_{1i}| B \rangle\ , \nonumber \\
{\rm or}\;\; \langle P V |O_{i}| B \rangle =& \langle P  | J_{4i}| 0 \rangle\ \langle V  | J_{3i}| B \rangle\ ,
\end{align}
where the $J_{ji}$ are the transition currents. This approach is known as {\it naive} factorization since it factorizes
 $\langle P V |O_{i}| B \rangle$  into a simple product of two quark matrix elements, (see Fig.~\ref{figch3.5}).
%
%
%~~~~~~~~~~~~~~~~~~~
%~~~~~~~~~~~~~~~~~~
%
Analytically, Fig.~\ref{figch3.5} can be written down as,
\begin{multline}\label{eq3.12a}
A(B \rightarrow P V)  \propto  \Biggl[ \sum_{i=1}^{10}  V_{CKM} C_{i}(\mu) \langle M_{1}M_{2}  | O_{i}
| B \rangle \Biggr] \\
 \propto  \Biggl[ \sum_{i=1}^{10}  V_{CKM}  C_{i}(\mu) \langle M_{1}  | J_{2i}| 0 \rangle
 \langle M_{2}  | J_{1i}| B \rangle  \Biggr]\ .
\end{multline}
A possible  justification for this approximation has been given by Bjorken~\cite{Bjorken:1989kk}:
the  heavy quark decays are very energetic, so the quark-antiquark pair in a meson in the  final state moves
very fast away from
the localised weak interaction. The hadronization of the quark-antiquark pair occurs far away from the remaining
 quarks. Then,
the meson can be factorized out and the interaction between the quark pair in the meson and the remaining quark
is  tiny.

The main uncertainty in this
approach is that  the  final state interactions (FSI)  are  neglected.   
 Corrections associated with the
factorization hypothesis are parameterized~\cite{ollie1, ollie2, ollie3}   
and hence there maybe    large uncertainties~\cite{Quinn:1999yq}. In spite
of this, there are indications that  should give 
a good estimate of the magnitude of the $B$ decay amplitude in many cases~\cite{Cheng:1994zx, Cheng:1997xy}. 
In order to improve  the estimate of the hadronic matrix element, we will briefly present in Section 2
 the formalism of QCD factorization, which is an extension of naive factorization.
We will see  how it is  possible to incorporate QCD corrections in order to include  the FSI at the first 
order in $\alpha_{s}$ into  the factorization approach. In Section 3, we will list our numerical results for the 
branching ratios related to the channels $B \to \rho \pi$ and $B \to \omega \pi$. In Section 4, we 
 will constrain the form factor $F_{1}^{B \to \pi}$ and propose  boundaries for this form factor depending on the 
CKM matrix element parameters $\rho$ and $\eta$. Finally, in the last section  we will summarize our analysis and draw
some conclusions.

%#########################################################################################
%-----------------------------------------------------------------------------------------
\section{QCD Factorization}
\begin{figure}[htpb]
\centering\includegraphics[height=3.6cm,clip=true]{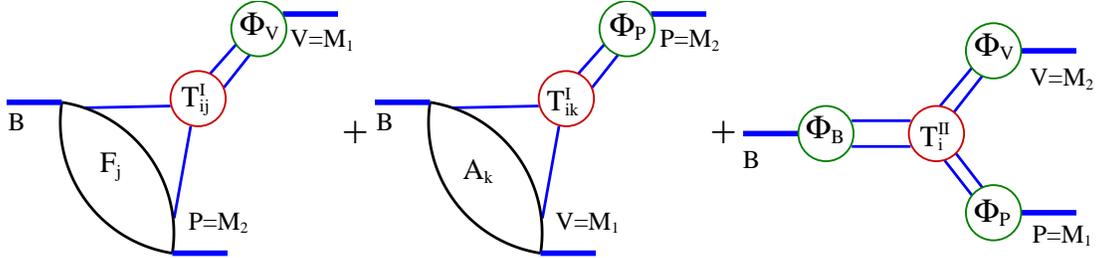}
\caption{Graphical representation of the QCD factorization formula.}
\label{fig10.1}
\end{figure}

%-----------------------------------------------------------------------------------------
%#########################################################################################
Factorization in charmless $B$ decays involves three fundamental scales: the weak interaction scale $M_{W}$,
 the $b$ quark mass scale $m_{b}$, and the strong interaction scale $\Lambda_{QCD}$. It is well known that the 
non-leptonic decay  amplitude for  $B \to P V$ is proportional to:
\begin{eqnarray}\label{eq10.1}
A(B \to P V) \propto \sum_{i} C_{i}(\mu) \langle P V | O_{i}(\mu) | B \rangle\ ,
\end{eqnarray}
where  we have omitted the CKM factor and Fermi constant for simplicity. The matrix elements $\langle P V | O_{i}(\mu) 
| B \rangle $  contain  non-perturbative effects which cannot be  accurately evaluated.
The coefficients $C_{i}(\mu)$ include  strong interaction effects from the scales $M_{W}$ down to $m_{b}$  and
is under   control. The aim is therefore  to obtain a good  estimate of the matrix 
elements without assuming naive factorization. 
In QCD factorization (QCDF), assuming a heavy quark expansion when $m_{b} \gg \Lambda_{QCD}$ and soft collinear 
factorization where the particle energies are bigger than the scale $\Lambda_{QCD}$, the matrix
elements  $\langle P V | O_{i}(\mu) | B \rangle$  can  be written as~\cite{Beneke:1999br}:
\begin{eqnarray}\label{eq10.2a}
\langle  PV | O_{i}(\mu) | B \rangle = \langle  P | j_{1} | B \rangle \langle  V | j_{2} | 0 \rangle
\biggl[ 1 + \sum_{n} r_{n} \alpha_{s}^{n} + \mathcal{O}(\Lambda_{QCD}/m_{b})\biggr]\ , 
\end{eqnarray}
where $r_{n}$ refers to the radiative corrections in $\alpha_{s}$ and $j_{i}$ are the quark currents.
It is straightforward to see that  if we neglect the corrections at the order $\alpha_{s}$, we recover the conventional
naive factorization in the heavy quark limit. 
We can rewrite the matrix elements  $\langle P V | O_{i}(\mu) | B \rangle$,
at the  leading order in $\Lambda_{QCD}/m_{b}$, in the QCDF approach by using a partonic language and one 
has~\cite{Beneke:1999br, Neubert:2001cq, Neubert:2000kk, Beneke:2000pw, Beneke:2002nj, Beneke:1999gt}:
\begin{multline}\label{eq10.2}
\langle  PV | O_{i}(\mu) | B \rangle = F_{j}^{B \rightarrow P}(m^{2}_{V}) \int_{0}^{1} d x T_{ij}^{I}(x) \phi_{V}(x)
+ A_{k}^{B \rightarrow V}(m^{2}_{P}) \int_{0}^{1} d y  T_{ik}^{I}(y) \phi_{p}(y) \\
+ \int_{0}^{1} d \xi  \int_{0}^{1} d x  \int_{0}^{1} d y T_{i}^{II}(\xi,x,y) \phi_{B}(\xi) \phi_{V}(x) \phi_{P}(y)\ ,
\end{multline}
where $\phi_{M}$ (with $M=V,P,B$) are the leading twist light cone distribution amplitudes (LCDA)
of valence quark Fock states. The light cone  momentum fractions of 
the constituent quarks of the vector, pseudoscalar and $B$ mesons are given respectively by $x,y,$ and $\xi$.
The form factors for $B \to P$ and $B \to V$ semi-leptonic decays evaluated at $k^{2}=0$ are denoted by 
$F_{j}^{B \rightarrow P}(m^{2}_{V})$ and $A_{k}^{B \rightarrow V}(m^{2}_{P})$.  Eq.~(\ref{eq10.2})
can be understood via Fig.~\ref{fig10.1} where a graphical representation of the factorization formula is given.
The hadronic decay amplitude involves both soft and hard contributions. At leading order, all 
the non-perturbative effects
are  contained in the  form factors  and the light cone distributions 
amplitudes. Then, non-factorizable interactions are dominated by hard gluon exchanges (in the 
case where the  $O(\Lambda_{QCD}/m_{b}$) terms are neglected) 
and can be  calculated  perturbatively, in order to correct the naive factorization approximation. 
 These hard scattering kernels~\cite{Beneke:1999br, Neubert:2001cq, Neubert:2000kk, Beneke:2000pw, Beneke:2002nj, 
Beneke:1999gt, Beneke:2000fw},   $T_{ik}^{I}$ and $T_{i}^{II}$, are calculable order
by order in perturbation theory. The naive factorization terms are recovered by the leading terms of $T_{ik}^{I}$
coming from the tree level  contributions, whereas  vertex corrections  and penguin 
corrections   are included at  higher orders of 
$\alpha_{s}$ in $T_{ik}^{I}$. The  hard interactions (at order  $O(\alpha_{s}$)) between the spectator quark 
and the emitted meson, at large  gluon momentum,  are taken into 
account  by $T_{i}^{II}$.
%
%

%
%~~~~~~~~~~~~~~~~~~~~~~~~~~~~~~~~~~~~~~~~~~~~~~~~~~~~~~~~~
\subsection{The QCD coefficients $a_{i}$}\label{sec10.2.1}
%~~~~~~~~~~~~~~~~~~~~~~~~~~~~~~~~~~~~~~~~~~~~~~~~~~~~~~~~~
%
%
The coefficients $a_{i}$~\cite{Beneke:2001ev, Beneke:2000ry}, 
have been calculated at 
next-to-leading order. They contain all the non-factorizable effects at order in $\alpha_{s}$. In order to clearly 
separate every contribution, the coefficients $a_{i}$  are written as the sum of, 
\begin{eqnarray}\label{eq10.6}
a_{i}= a_{i,I} +a_{i,II}\ ,
\end{eqnarray}
where the first term includes the naive factorization, the vertex and penguin corrections, while the second term 
contains the hard spectator interactions. According to the final states, the terms $a_{i}$ have to be expressed for 
two different cases: case A corresponds to the situation where the recoiling meson $M_{1}$  is a vector and the emitted
 meson $M_{2}$  is a pseudoscalar, and vice-versa for case B. For case A, the coefficients $a_{i}$ 
take the form~\cite{Beneke:2001ev, Beneke:2000ry}, 
%^{eff}
\begin{align}\label{eq10.6z}
a_{1,I} & = C_{1} + \frac{C_{2}}{N_{c}^{eff}} \Big[ 1 + \frac{C_{F}{\alpha}_{s}}{4{\pi}} V_{M}\Big]\ ,  &    
\!\!\! a_{1,II}&  = \frac{{\pi}C_{F}{\alpha}_{s}}{N_{c}^{eff 2}}C_{2} H(BM_{1},M_{2})\ ,  \nonumber\\
a_{2,I} & = C_{2} + \frac{C_{1}}{N_{c}^{eff}}\Big[1 + \frac{C_{F}{\alpha}_{s}}{4{\pi}} V_{M}\Big]\ , & 
\!\!\!a_{2,II}&  = \frac{{\pi}C_{F}{\alpha}_{s}}{N_{c}^{eff 2}} C_{1} H(BM_{1},M_{2})\ ,   \nonumber\\
a_{3,I} & = C_{3} + \frac{C_{4}}{N_{c}^{eff}}\Big[1 + \frac{C_{F}{\alpha}_{s}}{4{\pi}} V_{M}\Big]\ , & 
\!\!\!a_{3,II}&  = \frac{{\pi}C_{F}{\alpha}_{s}}{N_{c}^{eff 2}}C_{4} H(BM_{1},M_{2})\ ,  \nonumber \\
a_{4,I}^{p} & = C_{4} + \frac{C_{3}}{N_{c}^{eff}}\Big[1 + \frac{C_{F}{\alpha}_{s}}{4{\pi}} V_{M}\Big]
+a_{4,I,b}^{p}\ , & 
\!\!\!a_{4,II} & = \frac{{\pi}C_{F}{\alpha}_{s}}{N_{c}^{eff 2}}C_{3} H(BM_{1},M_{2})\ ,  \nonumber \\
a_{5,I} & = C_{5} + \frac{C_{6}}{N_{c}^{eff}}\Big[ 1 - \frac{C_{F}{\alpha}_{s}}{4{\pi}} V_{M}^{\prime} \Big]\ , & 
\!\!\!- a_{5,II} & =  \frac{{\pi}C_{F}{\alpha}_{s}}{N_{c}^{eff2}}C_{6} H^{\prime}(BM_{1},M_{2})\ ,   \nonumber\\
a_{6,I}^{p} & = C_{6} + \frac{C_{5}}{N_{c}^{eff}}\Big[ 1 - 6 \frac{C_{F}{\alpha}_{s}}{4{\pi}} \Big]
+ a_{6,I,b}^{p}\ , & 
\!\!\!a_{6,II} & = 0\ ,  \nonumber \\
a_{7,I} &  = C_{7} + \frac{C_{8}}{N_{c}^{eff}} \Big[ 1 - \frac{C_{F}{\alpha}_{s}}{4{\pi}} V_{M}^{\prime}  \Big]\ , & 
\!\!\!- a_{7,II} & =  \frac{{\pi}C_{F}{\alpha}_{s}}{N_{c}^{eff 2}} C_{8} H^{\prime}(BM_{1},M_{2})\ ,   \nonumber\\
a_{8,I}^{p} & = C_{8} + \frac{C_{7}}{N_{c}^{eff}} \Big[ 1 - 6 \frac{C_{F}{\alpha}_{s}}{4{\pi}}   \Big] 
+ a_{8,I,b}^{p}\ , & 
\!\!\!a_{8,II} & = 0\ ,  \nonumber \\
a_{9,I} & = C_{9} + \frac{C_{10}}{N_{c}^{eff}} \Big[     1 + \frac{C_{F}{\alpha}_{s}}{4{\pi}} V_{M}     \Big]\ , & 
\!\!\!a_{9,II} & = \frac{{\pi}C_{F}{\alpha}_{s}}{N_{c}^{eff 2}} C_{10} H(BM_{1},M_{2})\ ,   \nonumber\\
a_{10,I}^{p} & = C_{10} + \frac{C_{9}}{N_{c}^{eff}}  \Big[1 + \frac{C_{F}{\alpha}_{s}}{4{\pi}} V_{M}\Big]
           +a_{10,I,b}^{p} \ , & 
\!\!\!a_{10,II} & = \frac{{\pi}C_{F}{\alpha}_{s}}{N_{c}^{eff 2}}C_{9} H(BM_{1},M_{2})\ , 
\end{align}
where the terms $a_{4,I,b}^{p}, a_{6,I,b}^{p}, a_{8,I,b}^{p}$ and $a_{10,I,b}^{p}$ are, 
\begin{align}\label{eq10.8}
a_{4,I,b}^{p} & =  \frac{C_{F}{\alpha}_{s}}{4{\pi}} \frac{P_{M,2}^{p}}{N_{c}^{eff}}\ , & 
a_{6,I,b}^{p} & = \frac{C_{F}{\alpha}_{s}}{4{\pi}}  \frac{P_{M,3}^{p}}{N_{c}^{eff}}\ , \nonumber\\
a_{8,I,b}^{p} & = \frac{\alpha}{9{\pi}}     \frac{P_{M,3}^{p,ew}}{N_{c}^{eff}}\ , &
a_{10,I,b}^{p} & =  \frac{\alpha}{9{\pi}} \frac{P_{M,2}^{p,ew}}{N_{c}^{eff}}\ .
\end{align}
In Eqs.~(\ref{eq10.6z}) and~(\ref{eq10.8}) $V_{M}, V_{M}^{\prime}$ represent the vertex corrections, 
$H, H^{\prime}$ describe hard gluon exchanges between the spectator quark in the $B$ meson and the emitted meson 
(pseudoscalar or vector). $P_{M,2}^{p}, P_{M,3}^{p}, P_{M,3}^{p,ew}, P_{M,2}^{p,ew}$ are the QCD penguin 
contributions and electroweak penguin contributions, respectively. These quantities 
 are a result of the convolution of hard scattering kernels  $G$, with meson distribution 
amplitudes,  $\Phi$. We refer the reader to Refs.~\cite{Beneke:1999br, 
Neubert:2001cq, Neubert:2000kk, Beneke:2000pw, Beneke:2002nj, Beneke:1999gt} for more details.
Other parameters are $C_{i}\equiv C_{i}(\mu)$ (in NDR),
 $\alpha_s \equiv \alpha_{s}(\mu)$ (next to leading order), and $C_{F}= (N_{c}^{2}-1)/2 N_{c}$ with $N_{c}=3$. 
%The vertex corrections  $V_{M}$ and $V_{M}^{\prime}$ involved in $a_{i,I}$ are 
%given by~\cite{Beneke:2001ev, Beneke:2000ry}, 
%
%
%#########################################################################################
%-----------------------------------------------------------------------------------------
\section{Numerical results}
%-----------------------------------------------------------------------------------------
%#########################################################################################

\begin{figure}[htbp]
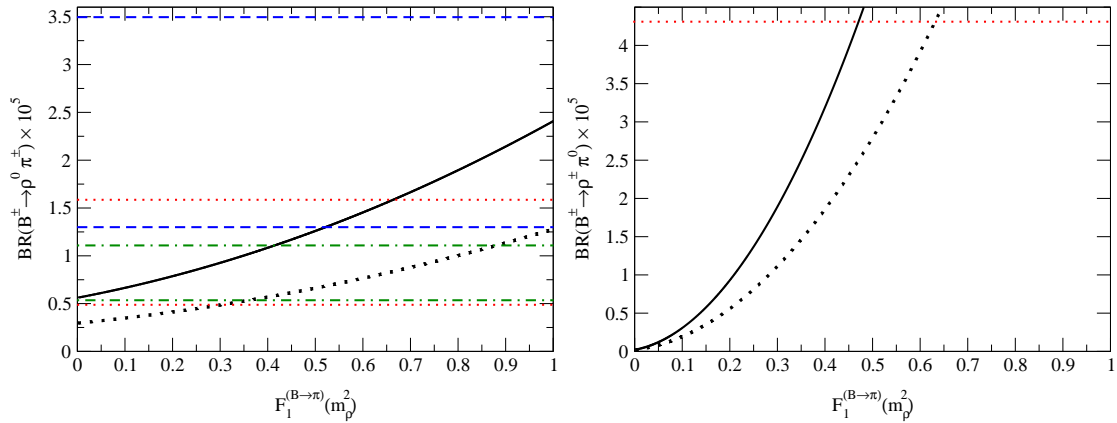

\includegraphics*[width=0.495\columnwidth]{QCD_frame_BR_rhoz_pip_th_talk.eps}
\includegraphics*[width=0.495\columnwidth]{QCD_frame_BR_rhop_pi0_th_talk.eps}
\caption{Branching ratio for $B^{\pm} \rightarrow \rho^{0}  \pi^{\pm}$, for  
limiting values of the  CKM matrix elements (Left hand-side). Branching ratio for $B^{\pm} 
\rightarrow \rho^{\pm}\pi^{0}$, for limiting values of the  CKM matrix elements (Right hand-side).
Solid line (dotted line) for  max (min) CKM matrix
elements. Notation: horizontal dotted  lines: CLEO data; horizontal dashed lines: BABAR data; horizontal 
dot-dashed lines: BELLE data.}
\label{graph11.12}
\end{figure}

\begin{figure}[htbp]
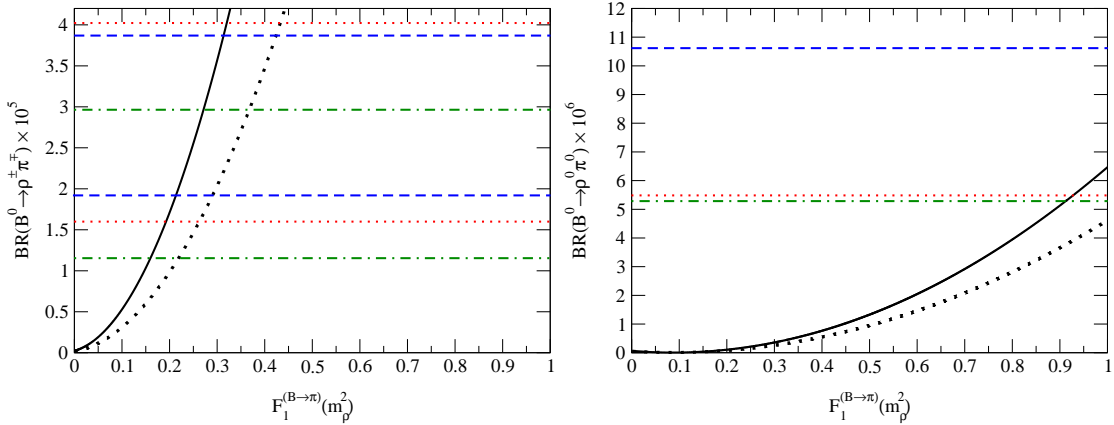

\includegraphics*[width=0.495\columnwidth]{QCD_frame_BR_rhop_pim_th_talk.eps}
\includegraphics*[width=0.495\columnwidth]{QCD_frame_BR_rho0_pi0_th_talk.eps}
\caption{ Branching ratio for $B^{0} \rightarrow \rho^{\pm}\pi^{\mp}$, for limiting 
values of the  CKM matrix elements (Left hand-side). Branching ratio for 
$B^{0} \rightarrow \rho^{0}\pi^{0}$, for limiting values of the  CKM matrix 
elements (Right hand-side). Solid line (dotted line) for  max (min) CKM matrix
elements. Notation: horizontal dotted  lines: CLEO data; horizontal dashed lines: BABAR data; horizontal 
dot-dashed lines: BELLE data.}
\label{graph11.34}
\end{figure}

\begin{figure}[htbp]
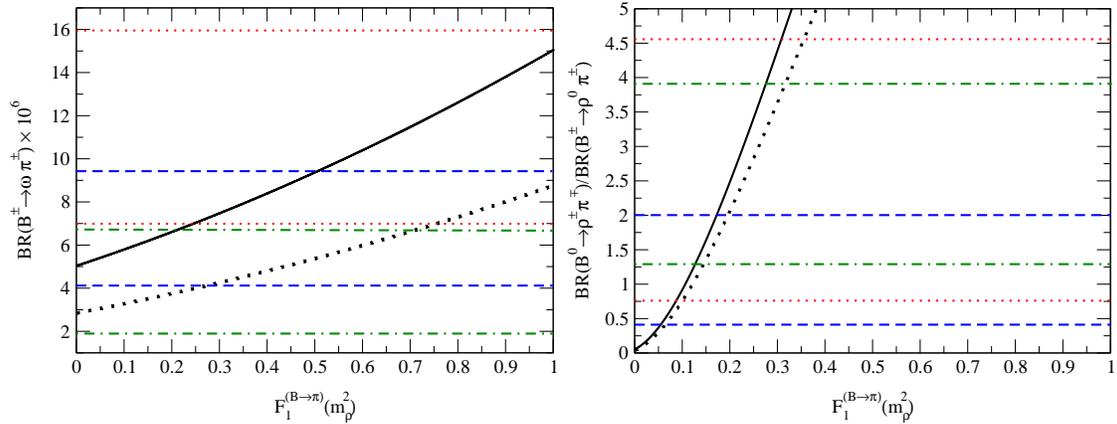

\includegraphics*[width=0.495\columnwidth]{QCD_frame_BR_omega_pip_th_talk.eps}
\includegraphics*[width=0.495\columnwidth]{QCD_frame_ratio_rhop_pm_rho0_pp_th_talk.eps}
\caption{ Branching ratio for $B^{\pm} \rightarrow \omega \pi^{\pm}$, for  
limiting values of the  CKM matrix 
elements (Left hand-side) .  The ratio of two $\rho \pi$ branching ratios limiting values of the CKM matrix
 elements (Right hand-side).
 Solid line (dotted line) for  max (min) CKM matrix
elements. Notation: horizontal dotted  lines: CLEO data; horizontal dashed lines: BABAR data; horizontal 
dot-dashed lines: BELLE data.}
\label{graph11.56}
\end{figure}

Assuming that all of the parameters involved in QCD factorization are constrained by independent
 studies where the 
 input parameters related to factorization were fitted, we concentrate our efforts on the form factor $F_{1}^{B\to
\pi}$ depending on  the CKM matrix parameters $\rho$ and $\eta$. In order to reach this aim, we have 
calculated the branching
ratios for $B$ decays such as $B^{\pm} \to \rho^{0} \pi^{\pm}, B^{0} \to \rho^{\pm} \pi^{0}, 
B^{0} \to \rho^{\pm} \pi^{\mp}, 
B^{0} \to \rho^{0} \pi^{0}$ and $B^{\pm} \to \omega  \pi^{\pm}$ where the annihilation and $\rho-\omega$ mixing
contributions were  taken into account. All the results are shown in 
Figs.~\ref{graph11.12},~\ref{graph11.34}  and~\ref{graph11.56}, 
and the  branching ratios are plotted as a function of the form factor $F_{1}^{B\to \pi}$ and as a function of 
 the values of $\rho$ and $\eta$ as well.

By taking into account experimental data from CLEO~\cite{Gao:1999ik, Jessop:2000bv, Schwarthoff:2002cf, 
DeMonchenault:2003pu, Briere:2002wq, Zhao:2001mz}, BELLE~\cite{Abe:2001pg, 
Bozek:2001xd, Lu:2002qp, Casey:2002yd, Abe:2002av, Garmash:2002yp, Gordon:2002yt, Iijima:2001gz, Kinoshita:2000uf}
and BABAR~\cite{Aubert:2001zf, Aubert:2002nc, Olsen:2000er, 
Aubert:2000vr, Schietinger:2001xp, Sciolla:2000gi, Cavoto:2001xn, Aubert:2001hs},  and comparing
theoretical predictions with experimental results, we expect to obtain a constraint on the form factor 
$F_{1}^{B\to \pi}$ depending on  the CKM matrix element parameters $\rho$ and $\eta$. Because of  the
 accuracy of the data, we shall
mainly use the CLEO and BELLE data for our analysis rather than those from BABAR. We expect that our results
should  depend more on uncertainties coming
from the experimental data than those from the factorization approach (as opposed to  naive factorization)
applied to calculate hadronic matrix
element  $\langle \rho \pi \vert J_{\mu} \vert B \rangle$ since in $B$ decays, $1/m_{b}$ corrections are very small.

%~~~~~~~~~~~~~~~~~~~
%~~~~~~~~~~~~~~~~~~~

For the branching ratio $B^{\pm} \to \rho^{0} \pi^{\pm}$ (Fig.~\ref{graph11.12}), we found   total consistency
between the theoretical results and experimental data from CLEO and BELLE. However, these results  allow us
to determine an upper  limit (between 0.40 and 0.65) for the value of the form factor $F_{1}^{B\to \pi}$. 
The weak dependence of the branching ratio on the form factor, $F_{1}^{B\to \pi}$,  is related to 
the tree and penguin amplitudes which are mainly governed by the form factor $A_{0}^{B\to \rho}$ rather than 
 $F_{1}^{B\to \pi}$.
Therefore, this branching ratio cannot be used as an efficient way to constrain the form factor $F_{1}^{B\to \pi}$.
 Note also that the comparison with BABAR data shows   agreement between theory and experiment when 
  $F_{1}^{B\to \pi}$ is bigger than 0.5.

For the branching ratio $B^{\pm} \to \rho^{\pm} \pi^{0}$ (Fig.~\ref{graph11.12}), CLEO gives only an upper
limit for the branching ratio  whereas BABAR and BELLE do not. Based on this upper limit, the value of the form
 factor $F_{1}^{B\to \pi}$ 
 must  be lower than 0.62. We emphasize that this  branching ratio is strongly dependent on the 
form factor $F_{1}^{B\to \pi}$ and hence provides an  efficient  constraint for  the value of   $F_{1}^{B\to \pi}$. 
For  the branching ratio
   $B^{0} \to \rho^{\pm} \pi^{\mp}$ (shown in Fig.~\ref{graph11.34}), BELLE, BABAR and CLEO give consistent
experimental data. The decay amplitude related to this branching ratio is  proportional to the form factor
$F_{1}^{B\to \pi}$ and thus  allows us to constrain the form factor effectively. Requiring  agreement between 
experimental values and 
theoretical results  yields a central  value for $F_{1}^{B\to \pi}$ which is about  0.3. Note  that for
 these  three
 branching ratios their  dependence  on the CKM matrix elements $\rho$ and $\eta$ 
is strong. Hence we  expect to be able to  determine   limits for their values when more $B$  decay channels are taken
into account.

%~~~~~~~~~~~~~~~~~~~
%~~~~~~~~~~~~~~~~~~

For the branching ratio $B^{0} \to \rho^{0} \pi^{0}$ (Fig.~\ref{graph11.34}), BABAR, BELLE and CLEO  only give 
an upper limit for the  branching ratio. However, the branching ratio does not appear  to be very
 sensitive to the CKM matrix elements $\rho$ and $\eta$. That could help us to obtain an upper limit for 
$F_{1}^{B\to \pi}$ which is not sensitive to $\rho$ and $\eta$. We therefore need new data to go further in this case.
Finally, we focus on the branching ratio $B^{\pm} \to \omega \pi^{\pm}$, plotted in Fig.~\ref{graph11.56}.
There is no agreement with the CLEO data for values of the  form factor $F_{1}^{B\to \pi}$ lower than 0.25  
whereas there is 
a good  agreement with BABAR and BELLE for any value of $F_{1}^{B\to \pi}$. Note 
that in this case the sensitivity of the branching ratio to the CKM matrix elements is bigger
than that to the form factor $F_{1}^{B\to \pi}$ and does not allow us to draw any conclusions regarding
the value of  $F_{1}^{B\to \pi}$.

To remove systematic errors in branching ratio data given by the $B$ factories, we can look at the ratio $R_{\pi}$ 
of the two following branching ratios: $\mathscr{B}(B^{0} \to \rho^{\pm} \pi^{\mp})$ and 
$\mathscr{B}(B^{\pm} \to \rho^{0} \pi^{\pm})$. In Fig.~\ref{graph11.56} we  show the ratio,  $R_{\pi}$, 
as a function of 
the form factor $F_{1}^{B\to \pi}$. All the $B$ factory data are in good agreement with theoretical predictions. 
The results indicate that the ratio is not   sensitive to the CKM matrix elements $\rho$ and $\eta$
whereas it is very sensitive to  the value of $F_{1}^{B\to \pi}$. Comparison with the  data shows that
   $F_{1}^{B\to \pi}$ is  
between 0.13 and 0.30 (BELLE),  0.05 and 0.20 (BABAR), and  0.10  and   0.35 (CLEO), respectively. Assuming that
 the value of $F_{1}^{B\to \pi}$  at $k^{2}=m_{\rho}^{2}$ is  around 0.30, we have
     $\mathscr{B}(B^{0} \to \rho^{\pm} \pi^{0}) 
\approx 14.2 \times 10^{-6}$ and $\mathscr{B}(B^{0} \to \rho^{0} \pi^{0}) < 1 \times 10^{-6}$. 

%
%~~~~~~~~~~~~~~~~~~~
%~~

It has to be pointed out that the annihilation
contributions in $B$ decays play an important role since they contribute significantly 
to the magnitude of the amplitude. The annihilation diagram contribution to the total decay  amplitude 
strongly modifies (in a positive or negative way) the branching ratio $B^{-} \rightarrow \rho^{0} \pi^{-}$ according
 to the value chosen for the phase $\phi_{A}$. This contribution could be  bigger than that of
 $\rho-\omega$ mixing  but carries more uncertainties because of its  endpoint divergence. 
 We emphasise that these two contributions ($\rho-\omega$ mixing effects and annihilation contributions) are not
just simple corrections to the total amplitude, but are important in  obtaining  a correct  description of
 $B$ decay amplitude.

%#########################################################################################
%-----------------------------------------------------------------------------------------
\section{Form factor $F_{1}^{B \to \pi}$}
%-----------------------------------------------------------------------------------------
%#########################################################################################

Form factors play a major role in the factorization method (naive or QCDF) since they represent  
the transition between two hadronic states.
Their computation is non trivial and may carry large uncertainties, depending on  models being used. These models 
include, say,  QCD sum rules, heavy quark effective theory, lattice QCD and  light cone QCD. 
With the available experimental data for the branching ratios, it is now possible for us to 
constrain $F_{1}^{B\to \pi}$ in a model-independent way in QCDF.

\begin{figure}[htbp]
\includegraphics*[width=0.595\columnwidth]{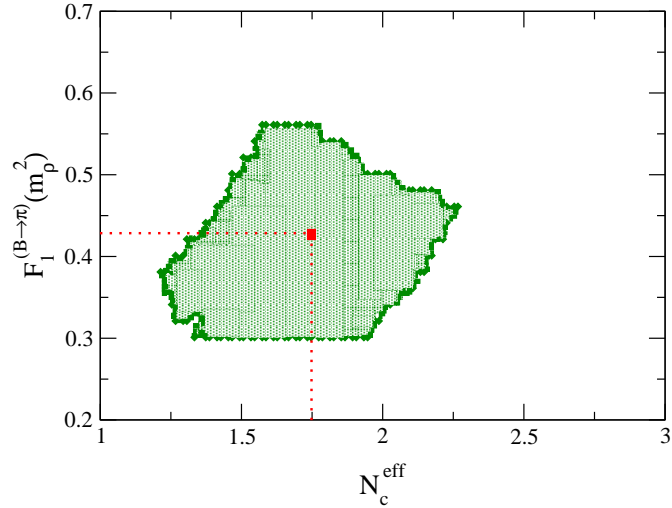}
\caption{$F_{1}^{B\to \pi}$ as a function of  $N_{c}^{eff}$. Plot  obtained by comparing theoretical results from 
QCFD with  experimental data from BELLE and CLEO for the 
 branching ratios  $B \to \rho \pi$ and  $B \to \omega \pi$. The plot includes  the uncertainties
from the CKM matrix element parameters  $\rho$ and $\eta$.}
\label{graph12.8}
\end{figure}

It has to be noticed  that the branching ratios depend on both $F_{1}^{B\to \pi}$ and $N_{c}^{eff}$.
In Fig.~\ref{graph12.8} we show  the results regarding the form factor 
 $F_{1}^{B \to \pi}(m_{\rho}^{2})$ as a function of $N_{c}^{eff}$, where we require that  
 all the branching ratios for  $B$ decaying into $\rho \pi$ and  $\omega \pi$ be consistent with the 
experimental data provided by  CLEO and BELLE. We have excluded the data from BABAR since they are currently 
not  numerous and accurate enough.
We have included  uncertainties from the CKM matrix element parameters $\rho$ ($0.190 < \rho < 0.268$) 
and $\eta$ ($0.284 < \eta < 0.366$) and we have 
applied the QCD factorization method where all of the final state interaction corrections  arising at  order
 $\alpha_s$ are incorporated. We emphasize that the  results are    model independent.

%~~~~~~~~~~~~~~~~~~~
%~~~~~~~~~~~~~~~~~~

We found a large common
region between  BELLE and CLEO for the $B$ decay into $\rho \pi$. From our analysis,
 $F_{1}^{B \to \pi}(m_{\rho}^{2})$ varies
between $0.3$ and $0.57$ and $N_{c}^{eff}$ can take  values from $1.25$ to $2.25$. Their central  values are
$F_{1}^{B \to \pi}(m_{\rho}^{2})=0.43$ and $N_{c}^{eff}=1.75$. 
The result obtained for the form factor $F_{1}^{B \to \pi}(m_{\rho}^{2})$ 
reduces one of the main uncertainties in the factorization process. That  obtained for the
effective number of colours, $N_{c}^{eff}$, confirms  previous analysis where naive 
factorization was applied for the same decays~\cite{ollie2}.

It is well known that the CKM matrix element parameters $\rho$ and $\eta$ are the main ``key'' to $CP$ violation 
within the Standard Model. Recall that the weak phase is mainly governed by the parameter $\eta$ that provides the 
imaginary part which is  absolutely necessary to obtain an asymmetry between matter and  antimatter.
Based on our analysis, we are not able  to   efficiently constrain the  CKM matrix parameters $\rho$ and $\eta$
   from the  branching ratios for $B \to \rho \pi$. In fact, the common region allowed by
CLEO and BELLE data for branching ratios for  $B \to \rho \pi$  does not constrain
the parameters $\rho$ and $\eta$. In the analysis  we used the values
$0.190 < \rho < 0.268$ and $ 0.284< \eta <0.366 $~\cite{Abbaneo:2001bv, Groom:2000in}, to which 
the common region corresponds. 
However, we can try (as an example) 
to get some constraints on $\rho$ and $\eta$ by only taking into account the central  values for the form factor
$F_{1}^{B \to \pi}(m_{\rho}^{2})$ and for   the effective number of colours 
$N_{c}^{eff}$.  According to our work, we find the following
limits: $0.205 <\rho<0.251$ and $0.300 < \eta<0.351$.

%#########################################################################################
%-----------------------------------------------------------------------------------------
\section{Conclusion}
%-----------------------------------------------------------------------------------------
%#########################################################################################

The calculation of the 
hadronic matrix elements that appear in the $B$ decay amplitude is non trivial. 
The main  difficulty is  to express the hadronic matrix elements
which represent the transition between the meson $B$ and the final state.

We  have investigated the branching ratios for $B \to \rho \pi,  B \to \omega \pi$ 
within the QCDF approach. Comparisons were made with experimental results from BABAR, BELLE and CLEO.
Based on our  analysis of branching ratios in $B$ decays, we have constrained the form factor, 
$F_{1}^{B \to \pi}(m_{\rho}^{2})$, and the effective number of colours,  $N_{c}^{eff}$. 
More accurate experimental data regarding branching ratios in $B$ decays  will  provide more accurate results, which  
will be helpful in  gaining  further  knowledge of direct $CP$ violation in $B$ decays.

This work could be extended to  more $B$ decays. It would be very interesting to 
constrain our parameters by investigating  channels other    than $\rho \pi$  for branching ratios and 
asymmetries. By including more channels, we will  use more experimental data and hence  be able to obtain 
better results for our parameters. In the QCD factorization  framework,
annihilation contributions could be subject to discussions. Clarifying this point would be very helpful in obtaining
 more accurate theoretical predictions. For example, it is important to solve the problem related to
the end point integral divergence~\cite{Beneke:1999br} which is parameterized without any 
strong physical motivation. Moreover, the annihilation contributions have not been included within the QCDF 
method. To obtain a consistent framework, it would be better to find a way   to include them within QCDF.
\newline
\newline
\noindent {\bf Acknowledgements}\\
This work was supported in part by the Australian Research Council and the
University of Adelaide.

%#########################################################################################


\begin{thebibliography}{99}
%
\bibitem{Buras:1999rb} Buras, Andrzej J., Lect. Notes Phys. {\bf 558} (2000) 65.
%
\bibitem{Buras:1998ra} Buras, Andrzej J., hep-ph/9806471.

\bibitem{Buchalla:1996vs}  Buchalla, Gerhard and Buras, Andrzej J. and Lautenbacher,
Markus E., Rev. Mod. Phys. {\bf 68} (1996) 1125.
%
\bibitem{Stech:1997ij} Stech, Berthold, hep-ph/9706384.
%
\bibitem{Buras:1995iy} Buras, Andrzej J., Nucl. Instrum. Meth. {\bf A368} (1995) 1.
%
\bibitem{Fakirov:1978ta} Fakirov, Dotcho and Stech, Berthold, Nucl. Phys. {\bf B133} (1978) 315.
%
\bibitem{Cabibbo:1978zv} Cabibbo, N. and Maiani, L., Phys. Lett. {\bf B73} (1978) 418.
%
\bibitem{Dugan:1991de} Dugan, Michael J. and Grinstein, Benjamin, Phys. Lett. {\bf B255} (1991) 583.
%
\bibitem{Bjorken:1989kk} Bjorken, James D., Nucl. Phys. Proc. Suppl. {\bf 11} (1989) 325.
%
%
\bibitem{ollie1} O. Leitner, X.-H. Guo and A.W. Thomas, Phys. Rev. {\bf D66} (2002) 096008.
%
\bibitem{ollie2} X.-H. Guo, O. Leitner and A.W. Thomas, Phys. Rev. {\bf D63} (2001) 056012.
%
\bibitem{ollie3} Z.J. Ajaltouni, O. Leitner, P. Perret, C. Rimbault and  A.W. Thomas,  
Eur. Phys. J. {\bf  C29} (2003), 215-233.
%
\bibitem{Quinn:1999yq} Quinn, Helen R., hep-ph/9912325.
%
\bibitem{Cheng:1994zx} Cheng, Hai-Yang, Phys. Lett. {\bf B335} (1994) 428.
%
\bibitem{Cheng:1997xy} Cheng, Hai-Yang, Phys. Lett. {\bf B395} (1997) 345.
%
\bibitem{Beneke:1999br} Beneke, M. and Buchalla, G. and Neubert, M. and Sachrajda,
 Christopher T., Phys. Rev. Lett. {\bf 83} (1999) 1914.
%
\bibitem{Neubert:2001cq} Neubert, Matthias, AIP Conf. Proc. {\bf 602} (2001) 168.
%
\bibitem{Neubert:2000kk} Neubert, Matthias, Nucl. Phys. Proc. Suppl. {\bf 99B} (2001) 113.
%
\bibitem{Beneke:2000pw} Beneke, M. J., Phys. {\bf G27} (2001) 1069.
%
\bibitem{Beneke:2002nj} Beneke, M., hep-ph/0207228.
%
\bibitem{Beneke:1999gt} Beneke, M., hep-ph/9910505. 
%
\bibitem{Beneke:2000fw} Beneke, M. and Buchalla, G. and Neubert, M. and Sachrajda,
Christopher T., hep-ph/0007256.
%
\bibitem{Beneke:2001ev} Beneke, M. and Buchalla, G. and Neubert, M. and Sachrajda,
Christopher T., Nucl. Phys. {\bf B606} (2001) 245.
%
\bibitem{Beneke:2000ry} Beneke, M. and Buchalla, G. and Neubert, M. and Sachrajda,
Christopher T., Nucl. Phys. {\bf B591} (2000) 313.
%
\bibitem{Gao:1999ik} Gao, Yongsheng and Wurthwein, Frank, CLEO collaboration, hep-ex/9904008. 
%
\bibitem{Jessop:2000bv} Jessop, C. P. and others, CLEO  collaboration, Phys. Rev. Lett. {\bf 85} (2000) 2881.
%
\bibitem{Schwarthoff:2002cf} Schwarthoff, H., CLEO  collaboration, hep-ex/0205015. 
%
\bibitem{DeMonchenault:2003pu} De Monchenault, Gautier Hamel, hep-ex/0305055.
%
\bibitem{Briere:2002wq} Briere, Roy A., CLEO  collaboration, AIP Conf. Proc. {\bf 618} (2002) 159.
%
\bibitem{Zhao:2001mz} Zhao, Xin, hep-ex/0101013.
%
\bibitem{Abe:2001pg} Abe, K. and others, BELLE collaboration, hep-ex/0107051.
%
\bibitem{Bozek:2001xd} Bozek, A., BELLE collaboration, hep-ex/0104041.
%
\bibitem{Lu:2002qp} Lu, R. S. and others, BELLE  collaboration, Phys. Rev. Lett. {\bf 89} (2002) 191801.
%
\bibitem{Casey:2002yd} Casey, B. C. K. and others, BELLE  collaboration, Phys. Rev. {\bf D66} (2002) 092002.
%
\bibitem{Abe:2002av} Abe, K. and others,  BELLE  collaboration, Phys. Rev. {\bf D65} (2002) 092005.
%
\bibitem{Garmash:2002yp} Garmash, Alexei, BELLE  collaboration, hep-ex/0207003.
%
\bibitem{Gordon:2002yt} Gordon, A. and others, BELLE  collaboration, Phys. Lett. {\bf B542} (2002) 183.
%)
\bibitem{Iijima:2001gz} Iijima, Toru, BELLE  collaboration, hep-ex/0105005.
%
\bibitem{Kinoshita:2000uf} Kinoshita, Kay, BELLE  collaboration, Nucl. Instrum. Meth. {\bf A462} (2001) 77.
%
\bibitem{Aubert:2001zf} Aubert, B. and others, BABAR  collaboration, Phys. Rev. Lett. {\bf 87} (2001) 221802.
%
\bibitem{Aubert:2002nc} Aubert, B. and others, BABAR  collaboration, hep-ex/0206004.
%
\bibitem{Olsen:2000er} Olsen, J., BABAR  collaboration, Int. J. Mod. Phys. {\bf A16S1A} (2001) 468.
%
\bibitem{Aubert:2000vr} Aubert, B. and others, BABAR  collaboration, hep-ex/0008058.
%
\bibitem{Schietinger:2001xp}  Schietinger, Thomas, BABAR  collaboration, hep-ex/0105019.
%
\bibitem{Sciolla:2000gi} Sciolla, G., BABAR  collaboration, Nucl. Phys. Proc. Suppl. {\bf 99B} (2001) 135.
%
\bibitem{Cavoto:2001xn} Cavoto, Gianluca, BABAR  collaboration, hep-ex/0105018.
%
\bibitem{Aubert:2001hs} Aubert, B. and others, BABAR  collaboration, Phys. Rev. Lett. {\bf 87} (2001) 151802.
%
\bibitem{Abbaneo:2001bv} Abbaneo, D. and others, hep-ex/0112028.
%
\bibitem{Groom:2000in} Groom, D. E. and others, Eur. Phys. J. {\bf  C15} (2000) 1.
%
\end{thebibliography}
\end{document}